# Free-space-coupled wavelength-scale disk resonators

Babak Mirzapourbeinekalaye, Sarath Samudrala, Mahdad Mansouree, Andrew McClung, and Amir Arbabi[1]

*Department of Electrical and Computer Engineering, University of Massachusetts Amherst, 151 Holdsworth Way, Amherst, MA 01003, USA*

Optical microresonators with low quality factor ($Q$) can be efficiently excited by and scatter freely propagating optical waves, but those with high $Q$ typically cannot. Here, we present a universal model for resonators interacting with freely propagating waves and show that the stored energy of a resonator excited by a plane wave is proportional to the product of its $Q$ and directivity. Guided by this result, we devise a microdisk with periodic protrusions in its circumference that couples efficiently to normally incident plane waves. We experimentally demonstrate several microdisk designs, including one with a radius of $0.75\lambda_0$ and $Q$ of 15,000. Our observation of thermally-induced bistability in this resonator at input powers as low as 0.7 mW confirms strong excitation. Their small footprints and mode volumes and the simplicity of their excitation and fabrication make wavelength-scale, free-space-coupled microdisks attractive for sensing, enhancing emission and nonlinearity, and as micro-laser cavities.

---

[1] arbabi@umass.edu

# Introduction

Optical microresonators are one of the main building blocks of photonic integrated circuits[1,2], lasers[3], and flat optical elements[4–6]. Waveguide-coupled on-chip microresonators such as microtoroid[7], microring, microdisk, and 1D and 2D photonic crystal resonators are used as filters[8,9], sensors[10,11], laser cavities[12], and for the enhancement of light-matter interaction[13], but they require coupling light to optical waveguides, which is an intricate and typically inefficient process. Free-space-coupled (FSC) optical resonators such as guided-mode resonators, vertical Fabry-Pérot resonators[14–16], and Mie resonators[17] can be readily excited and probed using free space excitations; however, they have larger footprints or lower quality factors ($Q$) than waveguide-coupled resonators. Although some high-$Q$ on-chip resonators can be probed using free-space illumination[18], they can only be weakly excited, and a crossed-polarization scheme is used to detect the small amount of light scattered by the resonant mode.

Large and dense arrays of FSC resonators such as Mie or vertical Fabry-Pérot resonators with subwavelength footprints and low and moderate $Q$s form flat optical elements such as metalenses[4,6]. However, higher $Q$ Mie resonators, which use high-order resonances cannot be efficiently excited using free-space excitations. High-$Q$ vertical Fabry-Pérot resonators that use distributed Bragg reflectors[19], such as the ones used in vertical-cavity surface-emitting lasers, have significantly larger footprints and mode sizes, and their resonant wavelengths cannot be controlled using lithographic patterning. High $Q$ and efficient free-space excitation are typical features of guided-mode resonances, also known as bound states in the continuum, but such resonant modes are not localized, and using a finite size resonator reduces their quality factors, thus limiting their minimum dimensions to tens of wavelengths[20–23]. Moreover, fabrication-induced nonuniformities cause further inhomogeneous broadening and reduction of their $Q$s.

Here we present the theory, design, and experimental demonstration of high-$Q$ resonators with a small footprint and mode size that can be efficiently excited and probed using free-space illumination. The resonators are subwavelength and wavelength-scale microdisks that are modified by adding weak azimuthal gratings that enable efficient free-space coupling. Azimuthal gratings have been previously added to microdisks and microrings to selectively increase or decrease their radiation quality factors[24,25], modify their dispersion and resonant wavelengths[26], generate beams carrying orbital angular momentums[27], selectively couple different degenerate resonant modes[28], and identify azimuthal orders of resonant modes[29,30]. In contrast, the main purpose of the azimuthal grating in this work is to efficiently couple incident plane waves to a resonant mode. The fabrication of the FSC microdisks requires only one lithography step, and their resonant wavelengths can be controlled lithographically. Because of their small footprints and localized modes, large and dense non-interacting arrays of these resonators can be realized. In the following, we introduce a universal model for the excitation of FSC resonators by plane waves and more general free propagating waves, and we present the operation principle and details of the design, fabrication, and characterization of FSC microdisk resonators.

## Results

Figure 1a schematically shows an arbitrary open resonator and the radiation pattern of its resonant mode. The mode has a resonant angular frequency $\omega_0$, a quality factor $Q$, and its radiation mode's directivity is $D(\theta, \phi)$. We assume that the resonator is excited by a plane wave with an angular frequency $\omega_0$ and energy density $u_i$. A coupled-mode equation that describes the dynamics of coupling between a high-$Q$ open resonator and an incident wave is presented in Supplementary Note 1, and expressions for the stored energy and absorbed power are given in Supplementary Note 2. As it is shown in Appendix II, the energy stored in a resonator made of lossless materials that is excited by the plane wave is given by

$$U_s = \frac{u_i \lambda_0^3}{2\pi^2} QD, \tag{1}$$

where $\lambda_0 = \frac{c}{2\pi\omega_0}$ is the free-space resonant wavelength of the resonator, and $c$ is the speed of light in a vacuum. The stored energy, which is an indicator of the excitation strength of the mode, is proportional to the product of $Q$ and the directivity $D$ of the resonators' radiation pattern along the incident direction of the plane wave. Therefore, a resonator can be efficiently excited by a plane wave if its $QD$ is large.

A microdisk resonator supports high-$Q$ modes, but its radiation pattern vanishes in the direction normal to the plane of the disk. Figure 1b shows a 0.25-μm-thick amorphous silicon (a-Si, $n = 3.6$) disk resonator with a diameter of 2.29 μm on a fused silica substrate. Figure 1b also shows the radiation pattern and the electric energy distribution of its resonant mode with a free-space wavelength of $\lambda_0 = 1.53$ μm computed using a commercial finite element solver[31]. The mode is quasi-transverse electric with a quality factor of $Q = 8.6 \times 10^4$, and its electric field radial component $E_\rho$ is proportional to $\cos N\phi$ for $N = 10$ ($\phi$: azimuthal coordinate). The directivity of the radiation pattern of the mode in the direction normal to the plane of the disk is zero; thus, the mode cannot be excited by a normally incident plane wave.

The polarization current density of the mode $\mathbf{J}^M = j\omega_0\epsilon_0(\epsilon_r - 1)\mathbf{E}^M$ is the source of radiation and can be examined for an intuitive understanding of the mode's radiation pattern ($\epsilon_0$: free-space permittivity, $\epsilon_r$: spatially dependent relative permittivity of the resonator, $\mathbf{E}^M$: electric field of the resonant mode, and superscript M standing for "mode"). The directivity along any direction is proportional to the modulus squared of the projection of $\mathbf{J}^M$ on the electric field of the incident wave[32]. The relative permittivity of the microdisk $\epsilon_r$ does not depend on $\phi$, thus $J_\rho^M$ varies as $\cos N\phi$, but the radial component of the electric field of an $x$-polarized plane wave normally incident on the microdisk varies as $\cos\phi$. Therefore, the projection of $\mathbf{J}^M$ on the incident electric field and the directivity along the normal direction are zero when $N \neq 1$. However, modifying $\epsilon_r$ by small amounts such that it contains terms that vary as $\cos(N \pm 1)\phi$ while the electric field of the mode is not significantly perturbed will lead to terms in $J_\rho^M$ that vary as $\cos\phi$ and thus nonzero directivity along the normal to the microdisk plane. Figure 1c shows a top view of the radial electric field of a resonant mode of a microdisk resonator with $N=5$. Slightly modifying the resonator by adding $N-1=4$ small perturbations around the microdisk (depicted by small circles in Fig. 1c) will lead to a term in $J_\rho^M$ that varies as $\cos\phi$ and can be considered an electric dipole in the microdisk plane. This dipole term couples to normally incident plane waves and effectively operates as a coupler between the plane wave and resonant mode of the microdisk.

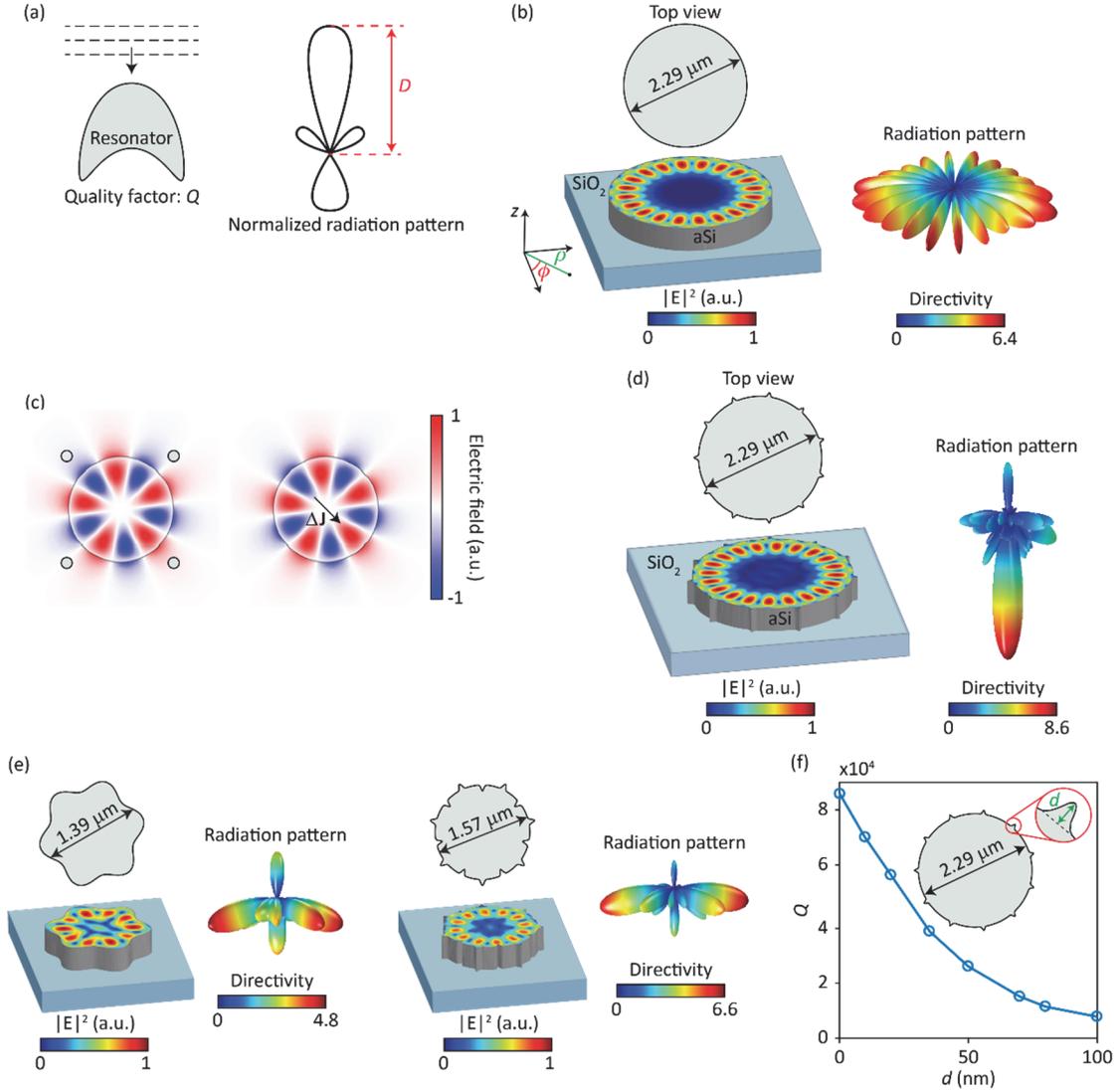

**Figure 1. Excitation of free-space-coupled resonators.** (a) Schematic illustration of a resonator with a quality factor $Q$ illuminated and excited with a plane wave. The directivity pattern $D(\theta,\phi)$ of the resonator, which is its normalized radiation pattern, is also shown. (b) Schematic of a microdisk resonator, a snapshot of the electric field squared of one of its resonant modes with azimuthal order $N=10$, and its radiation pattern. (c) The electric field of the resonant mode of a microdisk resonator. Small perturbations to the resonator are shown as small circles around the microdisk in the drawing on the left and can be represented by a polarization current density $\Delta \mathbf{J}$ as shown in the drawing on the right. (d) Resonant mode and radiation pattern of an FSC microdisk resonator for a mode with azimuthal order $N=10$. There are $N+1=11$ protrusions on the microdisk circumference. (e) Resonant modes and radiation patterns of two different FSC microdisk resonators for mode with azimuthal orders $N=5$ and 6. There are $N+1=6$ and 7 protrusions on the microdisks' circumferences. (f) The quality factor of the resonator shown in (d) as a function of the protrusion depth $d$.

The perturbations shown in Fig. 1c form an azimuthal grating. Azimuthal gratings could be realized by adding periodic structures close to the microdisk, but to reduce the device footprint, we chose to modulate its radius instead. The azimuthal gratings considered here are created by modifying the disk radius as $\rho = R_0 + d(\sin m\phi)^n$ for different values of $d, m,$ and $n$. The values of $m$ and $n$ were selected to have $N-1$ or $N+1$ grating periods around the microdisk circumference. The grating strength is a function of its maximum protrusion $d$ and its duty cycle, which the latter is controlled by $n$.

Figure 1d shows the resonant mode and the radiation pattern of the microdisk shown in Fig. 1b after adding $N+1 = 11$ protrusions. The radiation pattern has large lobes normal to the plane of the microdisk. Figure 1e shows smaller microdisks supporting modes with $N = 5$ and 6 at $\lambda_0 = 1.53$ μm and refractive index modifications with $N+1$ periods around the microdisk circumference. These resonators also emit significantly normal to their planes and can be excited by normally incident plane waves. The simulations confirm that weak azimuthal gratings with $N \pm 1$ periods around the microdisk circumference lead to emission of the resonant mode in the direction normal to the microdisk. The azimuthal grating serves a similar purpose as an evanescently coupled waveguide, but instead of coupling the resonant mode to a guided mode, it couples it to freely propagating waves. As the grating strength increases, the coupler loads the resonator more, and its quality factor decreases, as shown in Fig. 1f.

To confirm the excitation of the microdisk mode by normally incident plane waves and to quantify the effect of varying grating strengths, we found the energy stored in the microdisk when illuminated by a plane wave. Figure 2a shows a schematic of the simulated microdisk. The microdisk is the one shown in Fig. 1d and is illuminated by a plane wave normally incident from the substrate. The energy stored inside the microdisk is normalized to the energy of the incident plane wave in a $\lambda_0^3$ volume and is shown as a function of the radius modulation $d$ in Fig. 2b. As Fig. 2b shows, the stored energy is enhanced significantly by the addition of the grating, indicating the excitation of the resonant mode, and there is an optimal value for the grating strength ($d = 30$ nm for this resonator) that maximizes the stored energy and mode excitation amplitude. As mentioned earlier, the stored energy is proportional to the product of the directivity and the quality factor. The directivity increases and the quality factor decreases with increasing $d$; thus, there is an optimal $d$ value, or coupling strength, that maximizes their product. This behavior resembles the critical coupling condition in evanescently coupled resonators. Figure 2c shows the normalized stored energy versus wavelength for $d = 30$ nm. The narrow, Lorentzian line shape and its large peak value indicate efficient excitation of the resonant mode. The loaded $Q$ and the mode volume for this resonator are $Q = 4.33 \times 10^4$ and $V = 2.21 \left(\frac{\lambda_0}{n}\right)^3$ that result in a Purcell emission enhancement factor of $F = \frac{3}{4\pi^2} \frac{Q}{V/(\lambda_0/n)^3} = 1489$. The large Purcell factor and the directive radiation pattern of this resonator (Fig. 1d) make the FSC microdisks suitable for enhancing the rate of spontaneously emitted photons and efficiently collecting them.

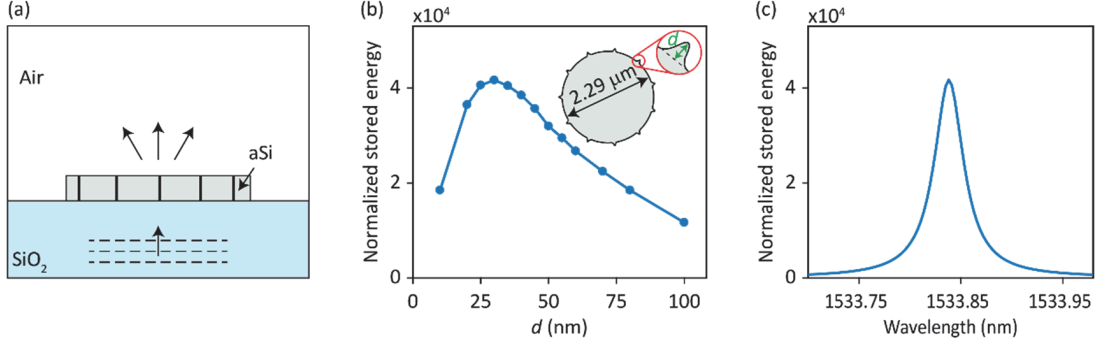

**Figure 2. Excitation of an FSC microdisk resonator by a plane wave.** (a) Illustration of an FSC microdisk resonator excited by a normally incident plane wave. (b) Normalized stored energy $U_s/(u_0\lambda_0^3)$ as a function of the grating protrusion $d$ at its resonant wavelength, and (c) as a function wavelength for $d = 30$ nm. The boundary of the microdisk is defined in polar coordinates according to $\rho = R_0 + d(\sin m\phi)^n$, where $R_0 = 1.145$ μm, $m = 5.5$, and $n = 100$.

To demonstrate the feasibility and performance of the FSC microdisk resonators, an array of microdisks were designed, fabricated, and characterized. Amorphous silicon microdisks with a thickness of 0.25 μm and diameters between 1.4 μm and 2.3 μm, corresponding to $N$ values from 5 to 10, were designed for operation at $\lambda_0 = 1.53$ μm. The microdisks were fabricated by depositing a 0.25-μm-thick layer of hydrogenated a-Si on a fused silica substrate, electron beam lithography using a negative resist, and dry etching (Methods). Figure 3a shows the SEM images of three fabricated microdisks with $N = 5$, 6, and 10 and $(d, m, n)$ values of (90 nm, 6, 1), (75 nm, 7, 21), and (30 nm, 5.5, 100), respectively.

The microdisks were characterized by measuring their transmission spectrum using the setup shown in Fig. 3b. The resonators were illuminated by focused polarized light from a tunable laser, and the transmitted light was collected using an objective lens and measured (Methods). Figure 3c shows an example of the measured spectrum for a microdisk with $N = 10$ (shown in Fig. 3a). The transmission spectrum typically shows a peak at the resonant wavelength of the device. However, depending on the alignment of the objective lenses, Fano line shapes (Fig. 3c inset) were also observed, which indicate interference between the light scattered by the microdisk resonant mode and unscattered light. For maximal excitation of the resonant mode, the numerical aperture of the excitation objective lens should be selected to match the angular divergence of the microdisk radiation pattern's lobe along the excitation direction (Supplementary Note 1). However, we did not optimize the objective lens and used a single lens for measuring different microdisks.

A microdisk supports two degenerate resonant modes at each resonant wavelength, both with $N$ field oscillations around its circumference whose difference is a rotation by half of the oscillation period (i.e., by angle $\pi/N$). The addition of the azimuthal grating with $N \pm 1$ periods does not break their degeneracy, but their coupler dipole moments (Fig. 1c) are rotated by 90°, and the two modes can be addressed independently using two orthogonal linear polarizations. Small and random variations in the microdisk radius created during the device fabrication remove the degeneracy of the two modes and split their resonant wavelengths[33]. Figure 3d shows two transmission spectra of a microdisk with $N = 10$ that are measured with two orthogonal linear polarizations. The splitting of the modes for this device is larger than the width of resonance peaks, and the two modes can be resolved.

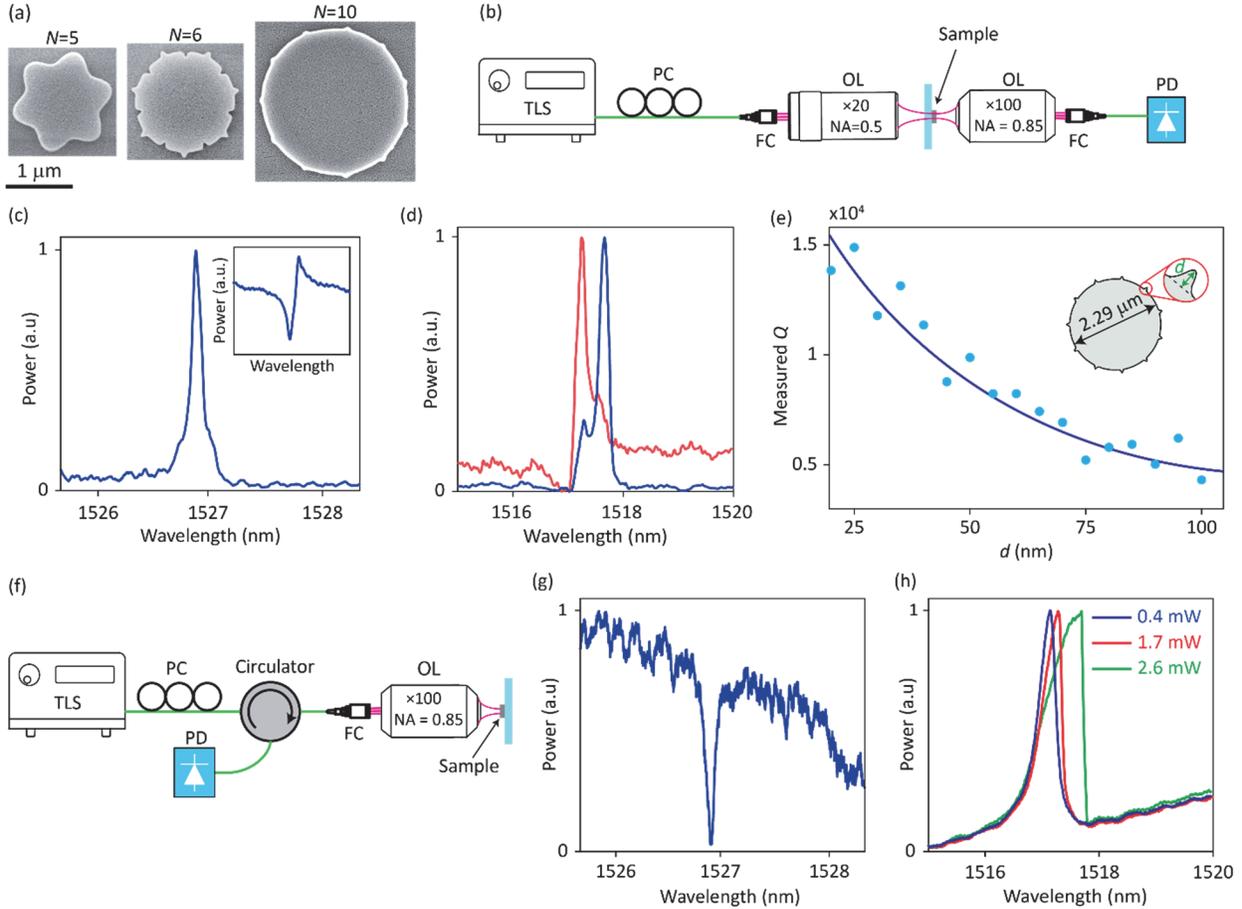

**Figure 3. Exprimental results.** (a) Scanning electron micrographs of fabricated FSC microdisk resonators with different azimuthal orders $N$. (b) Simplified schematic of the setup used for measuring transmission spectra of the resonators. (c) An example of a transmission spectrum of an FSC microdisk resonator. The inset shows another example of the observed transmission spectrum. (d) Measured transmission spectra of a resonator for two perpendicular linear polarizations of incident light. (e) Measured quality factors of FSC microdisks as a function of the protrusion depth $d$. The schematic of the resonator is shown in the inset. (f) Simplified schematic of the setup used for measuring reflection spectrum of FSC resonators. (g) Reflection spectrum of the resonator whose transmission spectrum is presented in (c). (h) Transmission spectra of an FSC resonator measured at different optical powers. TLS: tunable laser source, PC: polarization controller, OL: objective lens, FC: fiber coupler, PD: photodetector.

The quality factors of the microdisks were determined by fitting Fano line shapes to the measured transmission spectra (See Methods). Figure 3e shows the quality factor of microdisks with $N = 10$ as a function of the grating protrusion $d$. As expected, the quality factor decreases with increasing $d$; however, the highest measured quality factor was $1.5 \times 10^4$ which is by a factor a few smaller than simulated values (Fig. 1f). We attribute the reduction in $Q$ to absorption losses in a-Si and random radius variations and roughnesses of the fabricated microdisks. Replacing a-Si with crystalline silicon should reduce some of these losses and increase $Q$.

The microdisks can also be probed by measuring their reflection spectra using the confocal setup shown in Fig. 3f. Measuring the reflection spectrum is the preferred method for resonators fabricated on silicon-on-insulator (SOI) wafers. Figure 3g shows the reflection spectrum of the resonator, whose transmission spectrum is presented in Fig. 3c. A reflection dip is observed at the resonant wavelength of the microdisk, which is consistent with the peak observed in the transmission spectrum. Many resonators may be probed in series in reflection mode using the setup shown in Fig. 3f, or in parallel by imaging the light reflected from the resonators while tuning the incident light's wavelength.

To further verify the strong excitation of the resonant mode by free-space excitation, we measured the transmission spectrum of the microdisks at different incident optical powers. Thermally-induced optical nonlinearity, which is due to absorption losses in a-Si and its temperature-dependent refractive index, can be used as an indicator of the resonant mode's excitation strength. Figure 3h shows the transmission spectra of a microdisk with $d = 80$ nm, $m = 5.5$, and $n = 100$ obtained at different incident powers. Asymmetric line shapes, which are indicators of nonlinearities, were observed at sub-milliwatt incident power, and onsets of optical bistability (identified by the discontinuity in the measured transmission spectrum) as low as 0.7 mW were observed (Fig. S1). The observation of optical nonlinearities at such low incident powers confirms the strong excitation of the resonant mode.

## Discussion

The universal model we presented for the excitation of resonators by freely propagating waves describes the dynamic response of the resonator, provides simple relations for the accurate determination of stored energy and absorbed power, and enables the intuitive design of FSC resonators. FSC microdisk resonators with moderate and high $Q$ factors can be efficiently and readily excited and probed by free space illumination and collection (e.g., in a confocal arrangement). These resonators have subwavelength and wavelength-scale footprints enabling the realization of their dense arrays. Their small mode sizes and relatively high $Q$s lead to large enhancements of stored optical energies, nonlinearities, and emission rates (i.e., large Purcell factors), and their directional emission increases the collection efficiency. They can be fabricated using single-step lithography and a dry etch step. The devices presented here were implemented using a-Si on fused silica to enable the characterization of their transmission spectra. Similar resonators can be designed and implemented using SOI or silicon nitride on silicon dioxide platforms and achieve even higher $Q$ because of their smaller absorption losses. The small footprint and mode volume, simplicity of excitation and probing, moderate fabrication complexity, and the relatively high $Q$, make FSC microdisks attractive for sensing, filtering, enhancing emission and nonlinearity, and as cavities for micro-lasers.

## Methods

**Design.** An array of FSC microdisk resonators for six different azimuthal mode orders from $N = 5$ to 10 and for 17 different protrusion depths $d$ values from 20 nm to 100 nm in 5 nm steps were designed for the resonant wavelength of 1530 nm. The boundaries of the microdisks were defined in the polar coordinates according to as $\rho = R_0 + d(\sin m\phi)^n$.

**Fabrication.** To fabricate the FSC microdisks, a 250-nm-thick layer of hydrogenated a-Si was deposited on a fused silica substrate by plasma-enhanced chemical vapor deposition using silane at 190° C. A ~250-nm-thick layer of a negative electron beam resist (AR-N 7520.11 new, Allresist GmbH) was coated on the a-Si layer and baked at 90° C and then a layer of conductive polymer (AR-PC 5091, Allresist GmbH) was spin-coated on the resist and baked at 50° C to serve as a charge dissipation layer. The resonators' pattern was written on the resist using a 125-keV electron beam lithography system (ELS-F125, Elionix). Subsequently, the charge dissipation layer was removed using deionized water, and the resist was developed for 1 min in a developer (AR 300-47, Allresist GmbH). The resist's pattern was then transferred to the a-Si layer by inductively coupled reactive ion etching in a mixture of $SF_6$ and $C_4F_8$ gases, and the resist was removed using a solvent (Remover PG, Kayaku Advanced Materials Inc.).

**Characterization.** The transmission spectra of the FSC microdisk resonators were measured using the setup shown in Fig. S4. Polarized light from a tunable laser (AQ4312A, Ando) covering the 1480 nm-1580 nm range was amplified using an optical amplifier (FIBERAMP-BT 20, Photonetics), passed through a manual polarization controller (FPC560, Thorlabs), and was free-space coupled using a fiber collimation package (F240FC-C, Thorlabs). A fiber-coupled visible laser was also coupled to the same collimation package using an optical switch (EK703-FC, Thorlabs). The output beam was sampled using a beam splitter (BP108, Thorlabs) and the reflected power was monitored using a photodetector (PDA10CF, Thorlabs) and was used for the normalization of incident power. The beam transmitted through the beam splitter was focused through the substrate on a resonator using an objective lens (×20, 0.5NA, PE IR PlanAPO, Seiwa Optical). Light transmitted through and scattered by the resonator was collected by another objective lens (×100, 0.85NA, IR Plan, Nikon Hamamatsu) and was measured using a free-space-coupled photodiode (FGA01, Thorlabs) and imaged using an infrared camera (MicronViewer 7290A, Electrophysics). The resonator sample was mounted on a three-axis translation stage equipped with piezo actuators with submicron resolution (Picomotor Actuator, Newport Corporation) for accurate alignment of the resonators to the focused incident beam. The visible laser light reflected from the sample was imaged using the illumination objective lens and a tube lens (AC254-200-A, Thorlabs) on a visible camera (EO-5012M, Edmund Optics), assisting the alignment.

The reflection spectra of the resonators were also measured using the setup shown in Fig. S4 by exciting the resonator with a tunable laser through the ×100 objective lens. The same objective lens also collected the reflected light, and the reflected light was sent to a fiber-coupled photodetector (ETX 75, JDS Uniphase) using an optical fiber circulator (6015-3-FC, Thorlabs).

The quality factor values reported in Fig. 3e and Fig. S5b were obtained by fitting the measured resonant line shapes of the transmission spectra to $T(\omega) = \left| a/(\omega - \omega_c) + be^{j\phi} \right|^2$, where $\omega_c = \omega_0(1 + \frac{j}{2Q})$, and using real-valued parameters $a, b, \phi, \omega_0,$ and $Q$ as the fit parameters (Fig. S5a).

# Acknowledgments

This work was performed in part at the Center for Nanoscale Systems (CNS) at Harvard University, a member of the National Nanotechnology Coordinated Infrastructure Network (NNCI), which is supported by the National Science Foundation under NSF award no. 1541959. Additional device fabrication was performed in the Conte Nanotechnology Cleanroom at the University of Massachusetts Amherst.

# References

1. Yao, Z. *et al.* Integrated Silicon Photonic Microresonators: Emerging Technologies. *IEEE J. Sel. Top. Quantum Electron.* **24**, 1–24 (2018).
2. Wang, C. *et al.* Monolithic lithium niobate photonic circuits for Kerr frequency comb generation and modulation. *Nat. Commun.* **10**, 978 (2019).
3. Ellis, B. *et al.* Ultralow-threshold electrically pumped quantum-dot photonic-crystal nanocavity laser. *Nat. Photonics* **5**, 297–300 (2011).
4. Decker, M. *et al.* High-efficiency dielectric Huygens' surfaces. *Adv. Opt. Mater.* **3**, 813–820 (2015).
5. Yu, Y. F. *et al.* High-transmission dielectric metasurface with 2pi phase control at visible wavelengths. *Laser Photonics Rev.* **9**, 412–418 (2015).
6. Arbabi, A., Horie, Y., Ball, A. J., Bagheri, M. & Faraon, A. Subwavelength-thick lenses with high numerical apertures and large efficiency based on high-contrast transmitarrays. *Nat. Commun.* **6**, 7069 (2015).
7. Armani, D. K., Kippenberg, T. J., Spillane, S. M. & Vahala, K. J. Ultra-high-Q toroid microcavity on a chip. *Nature* **421**, 925–928 (2003).
8. Little, B. E., Chu, S. T., Haus, H. A., Foresi, J. & Laine, J.-P. Microring resonator channel dropping filters. *J. Light. Technol.* **15**, 998–1005 (1997).
9. Takano, H., Song, B.-S., Asano, T. & Noda, S. Highly efficient multi-channel drop filter in a two-dimensional hetero photonic crystal. *Opt. Express* **14**, 3491 (2006).
10. Chung-Yen Chao & Guo, L. J. Design and optimization of microring resonators in biochemical sensing applications. *J. Light. Technol.* **24**, 1395–1402 (2006).
11. Chow, E., Grot, A., Mirkarimi, L. W., Sigalas, M. & Girolami, G. Ultracompact biochemical sensor built with two-dimensional photonic crystal microcavity. *Opt. Lett.* **29**, 1093 (2004).
12. Painter, O. *et al.* Two-dimensional photonic band-Gap defect mode laser. *Science* **284**, 1819–21 (1999).
13. Lodahl, P., Mahmoodian, S. & Stobbe, S. Interfacing single photons and single quantum dots with photonic nanostructures. *Rev. Mod. Phys.* **87**, 347–400 (2015).
14. Walls, K. *et al.* Narrowband multispectral filter set for visible band. *Opt. Express* **20**, 21917–21923 (2012).
15. Horie, Y., Arbabi, A., Han, S. & Faraon, A. High resolution on-chip optical filter array based on double subwavelength grating reflectors. *Opt. Express* **23**, 29848–29854 (2015).
16. Horie, Y., Arbabi, A., Arbabi, E., Kamali, S. M. & Faraon, A. Wide bandwidth and high resolution planar filter array based on DBR-metasurface-DBR structures. *Opt. Express* **24**, 11677 (2016).

17. Kuznetsov, A. I., Miroshnichenko, A. E., Brongersma, M. L., Kivshar, Y. S. & Luk'yanchuk, B. Optically resonant dielectric nanostructures. *Science* **354**, (2016).

18. Englund, D. *et al.* Resonant excitation of a quantum dot strongly coupled to a photonic crystal nanocavity. *Phys. Rev. Lett.* **104**, 73904 (2010).

19. Reitzenstein, S. *et al.* AlAs∕GaAs micropillar cavities with quality factors exceeding 150.000. *Appl. Phys. Lett.* **90**, 251109 (2007).

20. Tibuleac, S. & Magnusson, R. Reflection and transmission guided-mode resonance filters. *J. Opt. Soc. Am. A* **14**, 1617–1626 (1997).

21. Taghizadeh, A., Mørk, J. & Chung, I.-S. Ultracompact resonator with high quality-factor based on a hybrid grating structure. *Opt. Express* **23**, 14913–14921 (2015).

22. Hsu, C. W., Zhen, B., Stone, A. D., Joannopoulos, J. D. & Soljačić, M. Bound states in the continuum. *Nat. Rev. Mater.* **1**, 16048 (2016).

23. Menon, S., Prosad, A., Krishna, A. S. L., Biswas, R. & Raghunathan, V. Resonant mode engineering in silicon compatible multilayer guided-mode resonance structures under Gaussian beam excitation condition. *J. Opt.* **23**, 105001 (2021).

24. Fujita, M. & Baba, T. Microgear laser. *Appl. Phys. Lett.* **80**, 2051–2053 (2002).

25. Arbabi, A., Kamali, S. M., Arbabi, E., Griffin, B. G. & Goddard, L. L. Grating integrated single mode microring laser. *Opt. Express* **23**, 5335 (2015).

26. Lee, J. Y. & Fauchet, P. M. Slow-light dispersion in periodically patterned silicon microring resonators. *Opt. Lett.* **37**, 58–60 (2012).

27. Cai, X. *et al.* Integrated Compact Optical Vortex Beam Emitters. *Science* **338**, 363–366 (2012).

28. Arbabi, A., Kang, Y. M., Lu, C.-Y. Y., Chow, E. & Goddard, L. L. Realization of a narrowband single wavelength microring mirror. *Appl. Phys. Lett.* **99**, 091105 (2011).

29. Arbabi, A. & Goddard, L. L. Measurements of the refractive indices and thermo-optic coefficients of Si3N4 and SiOx using microring resonances. *Opt. Lett.* **38**, 3878–3881 (2013).

30. Lu, X. *et al.* Chip-integrated visible–telecom entangled photon pair source for quantum communication. *Nat. Phys.* **15**, 373–381 (2019).

31. COMSOL Multiphysics.

32. Balanis, C. A. *Antenna theory: analysis and design*. (John Wiley & Sons, 2015).

33. Kippenberg, T. J., Spillane, S. M. & Vahala, K. J. Modal coupling in traveling-wave resonators. *Opt. Lett.* **27**, 1669–1671 (2002).

# Supplementary Information

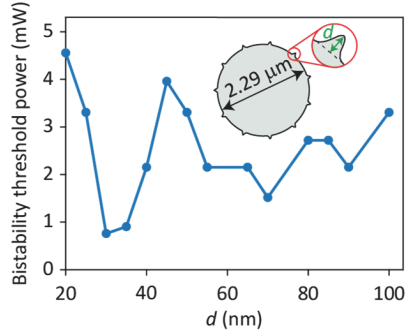

**Figure S1.** Measured minimum incident power for observing optical bistability for a set of resonators with different grating protrusion depth $d$. The boundary of the microdisk is defined in the polar coordinates according to $\rho = R_0 + d(\sin m\phi)^n$, where $R_0 = 1.145$ μm, $m = 5.5$, and $n = 100$.

## Supplementary Note 1: Temporal coupled-mode model for free-space-coupled resonators

Here we present a universal model for the excitation of resonators by freely propagating waves. The model describes the dynamic response of the resonator and can be used to determine the stored energy and the power absorbed in the resonator. Consider the resonator shown in Fig. S2a that is excited by a narrowband incident wave. For simplicity, we assume the electric field of the resonant mode $\mathbf{E}^M$ is normalized such that it radiates unit power. The phasor of the electric field of the resonant mode in the far-field can be written as

$$\mathbf{E}_{\mathrm{ff}} = \frac{-jk_0\eta_0}{\pi r} e^{-jk_0 r} \mathbf{F}(\theta, \phi), \tag{S1}$$

where $k_0 = \frac{\omega_0}{c}$ is the wavenumber at the resonant frequency $\omega_0$, $c$ is the speed of light in a vacuum, $\eta_0$ is the free-space impedance, and $\mathbf{F}(\theta, \phi)$ specifies the radiation pattern of the resonator. The directivity of the resonant mode pattern is given by[1]

$$D(\theta, \phi) = \frac{2\eta_0 k_0^2}{\pi} |\mathbf{F}(\theta, \phi)|^2. \tag{S2}$$

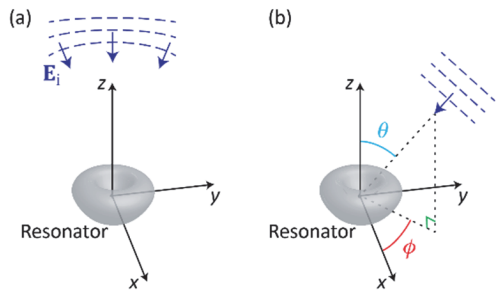

**Figure S2.** (a) Schematic illustration of a resonator excited by an incident wave. (b) Schematic illustration of a resonator excited by a plane wave incident from spherical coordinates direction $(\theta, \phi)$.

The dynamics of a driven resonator can be described using the time-domain coupled-mode model that describes the time evolution of the energy amplitude of the resonant mode[2]. The positive-frequency component of the electric field of the incident wave can be expressed as

$$\mathbf{E}_i = \mathcal{E}(t)e^{j\omega t}\tilde{\mathbf{E}}_i(\mathbf{r}), \tag{S3}$$

where $\mathcal{E}(t)$ is a slow varying complex-valued excitation amplitude, $\omega$ is the center frequency, and $\tilde{\mathbf{E}}_i(\mathbf{r})$ defines the spatial distribution of the incident wave.

The evolution of the energy amplitude of the resonant mode is described by

$$\frac{da}{dt} = j\omega_0 a - \frac{1}{2\tau}a + \kappa\mathcal{E}(t)e^{j\omega t}, \tag{S4}$$

where $a$ is the energy amplitude of the resonant mode (i.e., the stored energy of the resonant mode is given by $U_s = |a|^2$), $\omega_0$ is the resonant frequency, $\kappa$ is the coupling coefficient representing the coupling strength between the incident wave and the resonant mode, and $\tau = Q/\omega_0$ is the photon lifetime[2]. The photon loss rate is given by $1/\tau$, has contributions from the radiation and absorption losses, and can be written as $1/\tau = 1/\tau_r + 1/\tau_a$ where $\tau_r$ and $\tau_a$ are photon radiation and absorption lifetimes.

**Coupling coefficient for plane wave excitation**

We first find an expression for the coupling coefficient $\kappa$ for plane wave excitation and then determine $\kappa$ for more general excitations by expanding them in terms of plane waves. We assume that the resonator is excited by a narrowband plane wave that is incident from the spherical coordinates direction $(\theta, \phi)$, as shown in Fig. S2b. The positive-frequency component of the electric field of the incident wave can be expressed as

$$\mathbf{E}_i = \mathcal{E}(t)e^{j\omega t}e^{-j\mathbf{k}_i\cdot\mathbf{r}}\hat{e}, \tag{S5}$$

where $\mathbf{k}_i = -k_0(\sin\theta\cos\phi, \sin\theta\sin\phi, \cos\theta)$ is the wave vector, $k_0 = \frac{\omega}{c}$, and $\hat{e}$ is a unit vector defining the polarization. To find an expression for $\kappa$, we consider the special case when the resonator has no absorption lossless (i.e., $\tau = \tau_r$) and the incident wave is monochromatic with frequency $\omega_0$, that is $\mathcal{E}(t)e^{j\omega t} = E_0 e^{j\omega_0 t}$ where $E_0$ is a constant. Note that $\kappa$ is independent of absorption losses and $\mathcal{E}(t)$, and these assumption does not affect its value. Under such an excitation, $a = \tilde{a}e^{j\omega_0 t}$ and $\tilde{a}$ is found from (S4) as

$$\tilde{a} = 2\tau\kappa E_0. \tag{S6}$$

For a resonator with no absorption loss, the excitation amplitude of the resonant mode (the amplitude of $\mathbf{E}^M$ in the modal expansion of the scattered field) is given by[3,4]

$$b = \frac{1}{2}\int \mathbf{E}_i \cdot \mathbf{J}^M dv, \tag{S7}$$

where $\mathbf{J}^M$ is the equivalent polarization current density of the mode ($\mathbf{J}^M = j\omega_0\epsilon_0(\epsilon_r - 1)\mathbf{E}^M$). For a plane wave incident, the right hand of (S7) is a Fourier transform and is related to the radiation pattern of the current (which is the same as the mode) by[1]

$$b = 2E_0 \mathbf{F}(\theta, \phi) \cdot \hat{e}. \tag{S8}$$

Since we assumed $\mathbf{E}^M$ is power normalized, the power radiated by the excited resonant mode (i.e., $b\mathbf{E}^M$) is $|b|^2$. The radiated power is related to the stored energy of the mode $|\tilde{a}|^2$ by the radiative photon lifetime ($P_{\text{rad}} = U_s/\tau_r$), thus

$$b = \tilde{a}\sqrt{1/\tau_r}. \tag{S9}$$

From Eqs. (S6), (S8), and (S9), we find

$$\tilde{a} = 2\sqrt{\tau_r} E_0 \mathbf{F}(\theta, \phi) \cdot \hat{e} = 2\tau_r \kappa E_0, \tag{S10}$$

thus

$$\kappa = \sqrt{\frac{1}{\tau_r}} \mathbf{F}(\theta, \phi) \cdot \hat{e}. \tag{S11}$$

**Coupling coefficient for general excitation**

A more general coherent incident wave can be expanded in terms of plane waves, and the positive-frequency component of the electric field of the incident wave can be expressed as

$$\mathbf{E}_i = \mathcal{E}(t) e^{j\omega t} \tilde{\mathbf{E}}_i(\mathbf{r}) = \mathcal{E}(t) e^{j\omega t} \int \mathbf{e}(\theta, \phi) e^{-j\mathbf{k}_i \cdot \mathbf{r}} d\Omega, \tag{S12}$$

where $d\Omega = \sin\theta \, d\theta d\phi$ is the solid angle differential, and $\mathbf{e}(\theta, \phi)$ is the vector amplitude of the plane wave in the expansion of $\tilde{\mathbf{E}}_i(\mathbf{r})$ that is incident from the spherical coordinates' direction $(\theta, \phi)$. $\mathbf{e}(\theta, \phi)$ specifies the far-field radiation pattern of the incident wave. When $\mathbf{E}_i$ is incident from the $z > 0$ half-space, $\mathbf{e}$ is given by

$$\mathbf{e}(\theta, \phi) = k^2 \cos\theta \, \mathbb{E}_i(-k_0 \sin\theta \cos\phi, -k_0 \sin\theta \sin\phi), \tag{S13}$$

where $\mathbb{E}_i(k_x, k_y)$ is the spatial Fourier transform of $\tilde{\mathbf{E}}_i(\mathbf{r})$ in the $z = 0$ plane

$$\mathbb{E}_i(k_x, k_y) = \frac{1}{4\pi^2} \int \int \tilde{\mathbf{E}}_i(x, y, z=0) e^{j(k_x x + k_y y)} dx dy. \tag{S14}$$

The coupling coefficient for the general excitation is the sum of the coupling coefficients for different plane waves and is given by

$$\kappa = \sqrt{\frac{1}{\tau_r}} \int \mathbf{F}(\theta, \phi) \cdot \mathbf{e}(\theta, \phi) d\Omega. \tag{S15}$$

The time-averaged power of the incident wave is given by

$$P_{\text{in}} = |\mathcal{E}(t)|^2 \frac{2\pi^2}{\eta_0 k^2} \int |\mathbf{e}|^2 d\Omega. \tag{S16}$$

Based on (S15) and (S16) and according to the Cauchy–Schwarz inequality $\kappa$ is maximized when $\mathbf{e}(\theta, \phi)$ is proportional to $\mathbf{F}^*(\theta, \phi)$ and its maximum value is $\kappa_{\text{max}} = \sqrt{\frac{1}{\tau_r}} \frac{\sqrt{P_{\text{in}}}}{|\mathcal{E}(t)|}$. For a given incident power, the maximum stored energy is achieved when $\omega = \omega_0$, there is no absorption loss, and $\kappa$ has its maximum value. The maximum stored energy is given by $U_{s_{\text{max}}} = 4\tau_r P_{\text{in}}$. To increase coupling to the resonator, the overlap integral in (S15) should be maximized. For example, when exciting the resonator by a focused beam, the numerical aperture of the focusing lens should be selected such that the convergence of the incident beam matches the divergence of the radiation pattern of the resonator.

## Supplementary Note 2: Stored energy, reradiated and absorbed powers

Here we find the time-average stored energy, reradiated power, and absorbed power when the resonator is excited by a plane wave at its resonant frequency $\omega_0$. Plugging in $\kappa$ from (S11) in (S6), the phasor of the energy amplitude is found as

$$\tilde{a} = 2\tau \sqrt{\frac{1}{\tau_r}} \mathbf{F}(\theta, \phi) \cdot \hat{e} E_0. \tag{S17}$$

The stored energy

$$U_s = |\tilde{a}|^2 = \frac{4\pi\tau^2}{k_0^2 \tau_r} D \frac{1}{2\eta_0} |E_0|^2 = \frac{Q^2 D}{2\pi^2 Q_r} \left(\frac{1}{2}\epsilon_0 |E_0|^2 \lambda_0^3\right) = \frac{u_i \lambda_0^3}{2\pi^2} \frac{Q^2 D}{Q_r}, \tag{S18}$$

where we have used (S2) and have represented $D(\theta, \phi)$ by $D$ for brevity, and $u_i = \frac{1}{2}\epsilon_0|E_0|^2$ is the energy density of the incident wave. For a resonator with no absorption loss $Q = Q_r$ and (S18) reduces to (1). Figure S3 shows the normalized stored energy $U_s/(u_i \lambda_0^3)$ for the resonator shown in Fig. 1d computed using (1) and using direct integration of energy density obtained by full-wave simulations inside the resonator volume. The stored energy obtained using full-wave simulation contains small contributions from other non-resonant modes and is expected to be slightly larger.

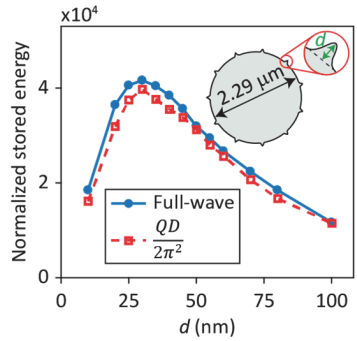

**Figure S3.** Normalized stored energy $U_s/(u_i \lambda_0^3)$ as a function of grating protrusion $d$ obtained using full-wave simulations and using (1). A schematic of the resonator used in the simulations is shown in the inset. The boundary of the microdisk is defined in the polar coordinates according to $\rho = R_0 + d(\sin m\phi)^n$, where $R_0 = 1.145$ μm, $m = 5.5$, and $n = 100$.

The resonator scatters some of the incident power, and a portion of the scattered power can be considered as the reradiation by the excited resonant mode. The reradiated power of the mode is given by

$$P_{rr}^M = \frac{U_s}{\tau_r} = 4\pi \frac{I_0}{k_0^2} \left(\frac{\tau}{\tau_r}\right)^2 D = 4A_{eff} I_0 \left(\frac{\tau}{\tau_r}\right)^2, \tag{S19}$$

where $I_0 = \frac{1}{2\eta_0}|E_0|^2$ is the intensity of the incident plane wave, and $A_{\text{eff}} = \frac{D\lambda_0^2}{4\pi}$ is the effective area of the resonator, which is the same expression used for the effective area of an antenna[1]. The scattering cross-section *due to the resonant mode* of the resonator is

$$\sigma_{\text{scat}}^{\text{M}} = 4A_{\text{eff}}\left(\frac{\tau}{\tau_r}\right)^2. \tag{S20}$$

$\sigma_{\text{scat}}^{\text{M}}$ is maximum for a resonator with no absorption loss and is equal to $4A_{\text{eff}}$. Note that the optical power scattered by the resonant mode might not be a significant portion of the total scattered power, and the total scattering cross-section of a high-$Q$ resonator might be significantly larger than $\sigma_{\text{scat}}^{\text{M}}$.

The power absorbed by the excited resonant mode is given by

$$P_a = \frac{U_s}{\tau_a} = U_s\left(\frac{1}{\tau} - \frac{1}{\tau_r}\right) = 4A_{\text{eff}}I_0\frac{\tau}{\tau_r}(1 - \frac{\tau}{\tau_r}), \tag{S21}$$

where we have used (S19) for $U_s$. For high-$Q$ resonators whose $A_{\text{eff}}$ are not very small, the power absorbed by the resonant mode is the dominant portion of total absorbed power, and the absorption cross-section of the resonator is given by

$$\sigma_{\text{abs}} = 4A_{\text{eff}}\frac{\tau}{\tau_r}(1 - \frac{\tau}{\tau_r}). \tag{S22}$$

The absorption cross-section is maximum when $\tau = 1/\tau_r$ or $\tau_r = \tau_a$ and its maximum value is $A_{\text{eff}}$. Therefore, the maximum power is absorbed by a resonator when its radiative and absorptive losses are equal, and we refer to such a resonator as a matched resonator. The maximum power that a resonator can absorb is

$$P_{a_{\max}} = A_{\text{eff}}I_0, \tag{S23}$$

which is the same as the maximum power that can be received by an impedance-matched antenna[1].

## References

1. Balanis, C. A. *Antenna theory: analysis and design.* (John Wiley & Sons, 2015).
2. Haus, H. *Waves and Fields in Optoelectronics. Prentice-Hall* (Prentice-Hall, 1985).
3. Harrington, R. & Mautz, J. Theory of characteristic modes for conducting bodies. *IEEE Trans. Antennas Propag.* **19**, 622–628 (1971).
4. Harrington, R., Mautz, J. & Yu Chang. Characteristic modes for dielectric and magnetic bodies. *IEEE Trans. Antennas Propag.* **20**, 194–198 (1972).

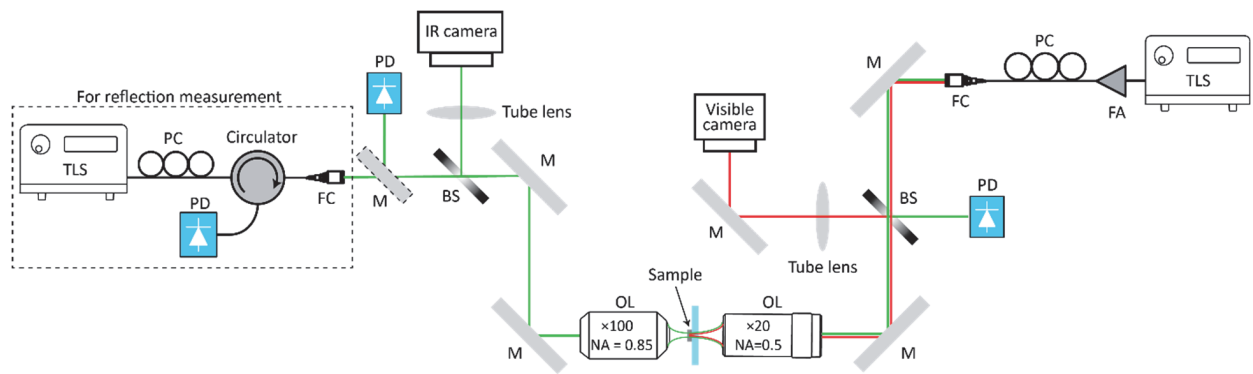

**Figure S4.** Schematic illustration of the characterization setup used for measuring transmission and reflection spectra and of microdisks at different power levels. PD: photodetector, M: mirror, BS: beam splitter, OL: objective lens, PC: polarization controller, FA: fiber amplifier FC: fiber collimator. The red and green lines depict the visible and infrared beam paths, respectively. The visible laser and camera are used for the imaging and alignment of the sample.

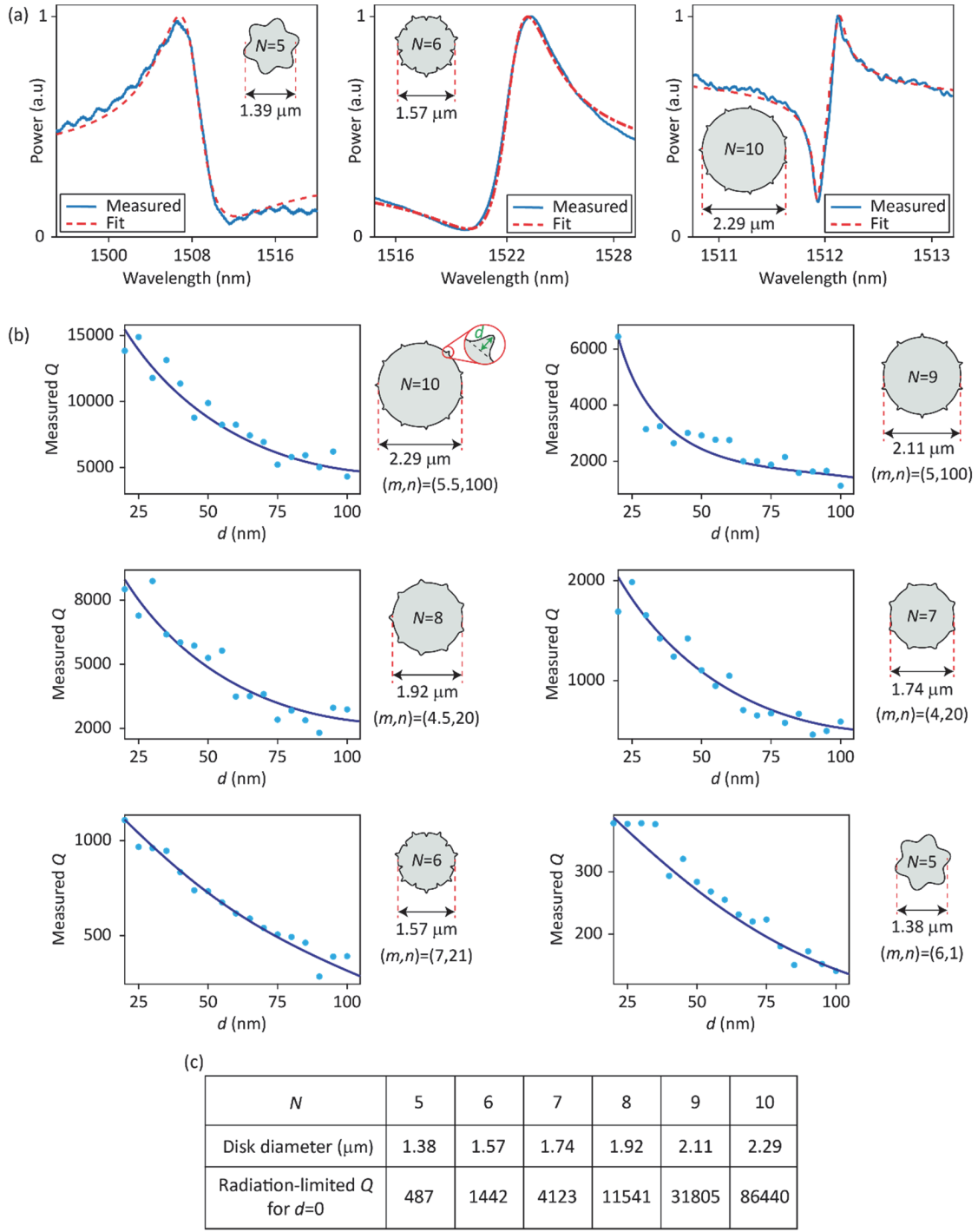

**Figure S5.** (a) Measured transmission spectra for three FSC microdisks representing Fano lineshapes. The boundary of the microdisk is defined in the polar coordinates according to $\rho = R_0 + d(\sin m\phi)^n$, where $R_0$ is the microdisk radius. The microdisks and their diameters are shown in insets, and $(d, m, n)$ are (80 nm, 6, 1), (40 nm, 7, 21), and (35 nm, 5.5, 100) for the three devices from left to right, respectively. Dashed lines are Fano lineshape fits. (b) Measured quality factors of an array of devices with different designs as functions of grating strengths $d$. (c) Simulated radiation-limited quality factors of the resonators presented in (b) with no protrusion ($d = 0$).

18